\documentclass[conference]{IEEEtran}
\IEEEoverridecommandlockouts
\usepackage{cite}
\usepackage{amsmath,amssymb,amsfonts}
\usepackage{algorithmic}
\usepackage{graphicx}
\usepackage{textcomp}
\usepackage{xcolor}
\def\BibTeX{{\rm B\kern-.05em{\sc i\kern-.025em b}\kern-.08em
    T\kern-.1667em\lower.7ex\hbox{E}\kern-.125emX}}
\begin{document}

\title{Machine Learning Techniques for Software Quality Assurance: A Survey\\
}

\author{\IEEEauthorblockN{Safa Omri}
\IEEEauthorblockA{\textit{Institute of Theretical Informatics} \\
\textit{Karlsruhe Institute of Technology}\\
Karlsruhe, Germany \\
safa.omri@kit.edu}
\and
\IEEEauthorblockN{Carsten Sinz}
\IEEEauthorblockA{\textit{Institute of Theretical Informatics} \\
\textit{Karlsruhe Institute of Technology}\\
Karlsruhe, Germany \\
carsten.sinz@kit.edu}

}

\maketitle

\begin{abstract}

%Software testing is an important and beneficial practice for improving software quality and reliability.
%It is a fundamental task to ensure software quality according to the specification of the system. 
%Regression testing necessary to validate the software after modification but it regarded as an expensive activity.
%Regression testing is one of the important software maintenance activities that let the software tester to ensure the quality and reliability of modified program. 

Over the last years, machine learning techniques have been applied to more and more application domains, including
software engineering and, especially, software quality assurance. Important application domains have been, e.g.,
software defect prediction or test case selection and prioritization.
The ability to predict which components in a large software system are most likely to contain the largest numbers of faults
in the next release helps to better manage projects, including early estimation of possible release delays, and affordably
guide corrective actions to improve the quality of the software. However, developing robust fault prediction models is a challenging
task and many techniques have been proposed in the literature.
Closely related to estimating defect-prone parts of a software system is the question of how to select and prioritize test cases, and indeed
test case prioritization has been extensively researched as a means for reducing the time taken to discover regressions in software.
In this survey, we discuss various approaches in both fault prediction and test case prioritization, also explaining how in recent studies deep learning algorithms for fault prediction help to bridge the gap between programs' semantics and fault prediction features.
We also review recently proposed machine learning methods for test case prioritization (TCP), and their ability to
reduce the cost of regression testing without negatively affecting fault detection capabilities.

\end{abstract}


\section{Introduction}

%Regression testing is one of the important software maintenance activities that let the software tester to ensure the quality and reliability of modified program. 
%However, as software evolves and becomes increasingly complex, the number of tests that are required to ensure correct functionality also grows, leading to a commensurate increase in the time taken to execute the test suite. 
%Since it often may take too much time to re-execute all of the tests for every change made to the system, developers may not know whether or not they have introduced regressions.

Nowadays, software quality assurance is overall the most expensive activity for nearly all software developing companies \cite{rana2014a}, since team members need to spend a significant amount of their time inspecting the entire software in detail rather than, for example, implementing new features. 
%If bugs are detected, the fixing of those consumes further development time.
Regression testing is one of the important software maintenance activities that let the software tester to ensure the quality and reliability of modified program. 
However, as software evolves and becomes increasingly complex, the number of tests that are required to ensure correct functionality also grows, leading to a commensurate increase in the time taken to execute the test suite. 
Since it often may take too much time to re-execute all of the tests for every change made to the system, it might be more efficient to first predict where which part of a software are most likely to be fault-prone and then prioritize the test cases in that part. If defect prediction can accurately predict the class that is most likely to be buggy, a tool can prioritize tests to rapidely detect the defects in that class. The more accurate a defect prediction is, the smaller the subset of the test suite needed to find potential bugs.
Therefore, software quality assurance activities, such as source code inspection, assist developers in finding potential bugs and allocating their testing efforts. They have a great influence on producing high quality reliable software.
Numerous research studies 
%\cite{hassan2009}, \cite{jiang2013}, \cite{xiao2014}, \cite{kim2007}, \cite{lee2011}, \cite{meneely2016}, \cite{moser2008}, \cite{nagappan2007}, \cite{rahman2013}, \cite{wang2012}, \cite{zimmermann2007} 
have analyzed software fault prediction techniques to help prioritize software testing and debugging. Software fault prediction is a process of building classifiers to anticipate which software modules or code areas are most likely to fail. Most of these techniques focus on designing features (e.g. complexity metrics) that correlate with potentially defective code. 
Object-oriented metrics were initially suggested by Chidamber and Kemerer \cite{chidamber1994a}. Basili et al. \cite{basili1996} and Briand et al. \cite{briand1999a} were among the first to use such metrics to validate and evaluate fault-proneness. Subramanyam and Krishnan \cite{subramanyam2003a} and Tang et al. \cite{tang1999a} showed that these metrics can be used as early indicators of externally visible software quality.
D'Ambros et al. have compared popular fault prediction approaches for software systems \cite{dambros2012}, namely, process metrics \cite{moser2008}, previous faults \cite{kim2007} and source code metrics \cite{basili1996}.
Nagappan et al. \cite{nagappan2006a} presented empirical evidence that code complexity metrics can predict post-release faults. 
Omri et al.'s
work \cite{omri2018} takes into consideration not only code complexity metrics but also the faults detected by static analysis tools to build accurate pre-release fault predictors.
Numerous research studies have analyzed code churn (number of lines of code added, removed, etc.) as a variable for predicting faults in large software systems \cite{khoshgoftaar1996, nagappan2005, omri2019}. 
All these research studies have gone into carefully designing features which are able to discriminate defective code from non-defective code such as code size, code complexity (e.g. Halstead, McCabe, CK features), code churn metrics (e.g. the number of code lines changed), or process metrics. 
Most defect prediction approaches consider defect prediction as a binary classification problem that can be solved by classification algorithms, e.g., Support Vector Machines
(SVM), Naive Bayes (NB), Decision Trees (DT), or Neural Networks (NN).
Such approaches simply classify source code changes into two categories: fault-prone or not fault-prone.

\begin{figure}[hbtp]
\centering
\includegraphics[scale=0.3]{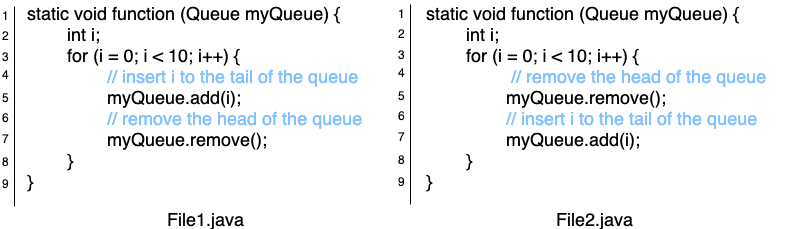}
\caption{A motivating example: \texttt{File2.java} will exhibit an exception when
the function is called with an empty queue.}
\label{fig:motivating-example}
\end{figure}

Those approaches, however, do not sufficiently capture the syntax and different levels of semantics of source code, which is an important capability for building accurate prediction models. 
Specifically, in order to make accurate predictions, features need to be discriminative: capable of distinguishing one instance of code region from another.
The existing traditional features cannot distinguish code regions with different semantics but
similar code structure.
For example, in Figure \ref{fig:motivating-example}, there are two Java files, both of which contain a \texttt{for} statement, a \texttt{remove} function and an \texttt{add} function. The only difference between the two files is the order of the \texttt{remove} and \texttt{add} function. \emph{File2.java} will produce a \texttt{NoSuchElementException} when the function is called with an empty queue.
%Treating program as bag of words without order, the state-of-the-art methods often overlook this local structural information.
Using traditional features to represent these two files, their feature vectors are identical, because these two files have the same source code characteristics in terms of lines of code, function calls, raw programming tokens, etc. However, the semantic content is different. Features that can distinguish such semantic differences should enable the building of more accurate prediction models.
%Natural Language Processing techniques have also been leveraged to extract defect predictors from code tokens in source files. A common technique is using Bag-of-Words (BoW) which treats code tokens as terms and represents a source file as term-frequencies. The BoW approach is however unable to detect differences in the se- mantics of source code due to differences in code order or syntactic structure (e.g. x ≥ y vs. y ≥ x ). Hence, recent trends started to focus on persevering code structure information in representing source code. However, recent work such as [28] does not fully encode the syntactic structure of code nor the semantics of code tokens, e.g. fails to recognize the semantic relations between “for” and “while”.
To bridge the gap between programs' semantic information and features used for defect prediction, some approaches propose to leverage a powerful representation-learning algorithm, namely deep learning, to capture the semantic representation of programs automatically and use this representation
to improve defect prediction. 
%Specifically, we use Deep Belief Network (DBN) [16] to automatically learn features from token vectors extracted from programs’ ASTs, and then utilize these features to train a defect prediction model.

After allocating the testing efforts based on fault prediction techniques, we need to know how much test cases needed to be executed within the predicted fault-prone components.
Within a software development process, software testing consumes a longer time in execution and can be the most expensive phase.
Researchers have developed a number of regression testing techniques \cite{yoo2012} to reduce the time required to detect regression. These techniques include test case selection, where information about the current set of changes (i.e., the classes or methods that were modified) is used to define a subset of tests that can be used to detect possible regressions. Test case minimization aims to find test cases that are redundant or irrelevant with respect to new tests that may perform similar actions. Finally, Test-case prioritization aims to determine the tests that are most likely to detect an error and to put them in the first priority in order to enable a quick detection of potential regressions. The use of test case prioritization (TCP) seems to improve the feasibility of tests in the software testing activity \cite{rothermel1999}.
Test case prioritization (TCP) techniques reduce the cost of testing using all test cases. Costs are reduced by parallelizing debugging and testing activities \cite{do2010}. The advantage of prioritizing test cases is that it enables continuous testing until resources are consumed or until all test cases are executed, although the most important test cases are always executed first \cite{rothermel2001}. The techniques for prioritizing test cases are classified into two groups: code-based and model-based techniques.  The code-based TCP techniques use the source code of the application to prioritize Test Cases. The model-based TCP techniques use the model that shows the required behavior of the system to prioritize Test Cases. Numerous techniques are proposed and developed for the prioritization of test cases. The most commonly used approaches are coverage-based \cite{nardo2015}, error-based, model-based, history-based, change-based, similarity-based, gene-based, etc. \cite{do2010}, \cite{goekce2014}, \cite{khatibsyarbini2017}.
However, traceability links between code and test cases are not always available or easily accessible when the test cases correspond to system tests. For example, in system testing, Test Cases are designed to test the whole system rather than simple units of code. Therefore, test case selection and prioritization must be handled differently, and the use of historical data on test case failures and successes has been suggested as an alternative \cite{kim2002}. Under the assumption that test cases that have failed in the past are more likely to fail in the future, these test cases are scheduled first in new regression testing cycles for history-based test case prioritization \cite{marijan2013}.
Since the time needed for the test cases varies strongly, not all tests can be executed, so that a test case selection is necessary. Although algorithms have recently been proposed \cite{marijan2013}, \cite{noor2015}, these two aspects of regression testing, the test-case selection and the history-based prioritization, can hardly be solved by using only non-adaptive methods. Non-adaptive methods may not be able to detect changes in the importance of some test cases compared to others because they use systematic prioritization algorithms.
%%%%%%
%%%%%%
%First, the time allocated to test case selection and prioritization in CI is limited as each step of the process is given a contract of time. So, time-effective methods shall be privileged over costly and complex prioritization algorithms. Second, history- based prioritization is not well adapted to changes in the execution environment. More precisely, it is frequent to see some test cases being removed from one cycle to another because they test an obsolete feature of the system. At the same time, new test cases are introduced to test new or changed features. Additionally, some test cases are more crucial in certain periods of time, because they test features on which customers focus the most, and then they loose their prevalence because the testing focus has changed. In brief, non-adaptive methods may not be able to spot changes in the importance of some test cases over others because they apply systematic prioritization algorithms.
%%%%reinforcement learning paper
To overcome these problems, Spieker et al. \cite{spieker2017} propose a new lightweight test case selection and prioritization approach to regression testing based on reinforcement learning and neural networks. Reinforcement Learning is well adapted to design an adaptive method that is able to learn from its experience with the execution environment.
This adaptive method can gradually improve its efficiency by monitoring the effects of its actions. By using a neural network that works with both the selected test cases and the order in which they are executed, the method tends to select and prioritize test cases that have been successfully used to detect errors in previous regression testing cycles and prioritize them so that the most promising ones are executed first.
The use of machine learning in TCP has become more and more interesting in the last years. By using learning algorithms the system can use the knowledge from previous tests and adapt to changing applications. Test cases define the priority with continuous actions and machine learning techniques offer easy adaptation to changes.

In this survey, we describe how defect prediction helps to obtain an early estimate of software component’s fault-proneness in order to guide software quality assurance towards inspecting and testing the components most likely to contain faults, with prioritizing the test cases. We review the different common machine learning techniques and the different deep learning technologies used in software quality assurance to predict faults, and provide a survey on the state-of-the-art in deep learning methods applied to software defect prediction. We also investigate some of the studies that use machine learning in the process of test case prioritization.
%We also dedicate complete sections on tackling safety aspects, the challenge of training data sources and the required computational hardware.
%Wouldn’t it be great if you could answer the classic testing question, “If I’ve made a change in this piece of code, what’s the minimum number of tests I should be able to run in order to figure out whether or not this change is good or bad?” Conceptually, if we can identify the parts of a system-under-test (SUT) affected by changes to a program, then we can prioritize on testing those parts and provide cost-effective testing solutions.

%This paper aims to explore the recent use of Machine Learning methods in the field of automated software testing, and particularly, regression testing, which is an ideal target for ML or autonomous testing. 

\section{Software Defect Prediction Process}
\begin{figure}[hbtp]
\centering
\includegraphics[scale=0.3]{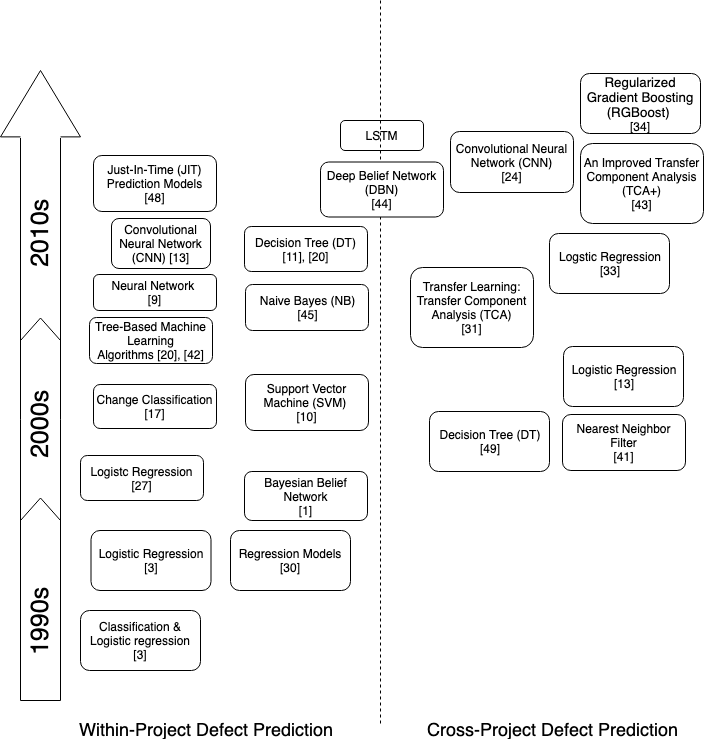}
\caption{History of Software Defect Prediction}
\label{fig:history-software-defect-prediction-studies}
\end{figure}

Fault prediction is an active research area in the field of software engineering. Many techniques and metrics have been developed to improve fault prediction performance.
In recent decades, numerous studies have examined the realm of software fault prediction. Figure \ref{fig:history-software-defect-prediction-studies} briefly shows the history of software fault prediction studies in about the last 20 years. 

\begin{figure*}[hbtp]
\centering
\includegraphics[scale=0.3]{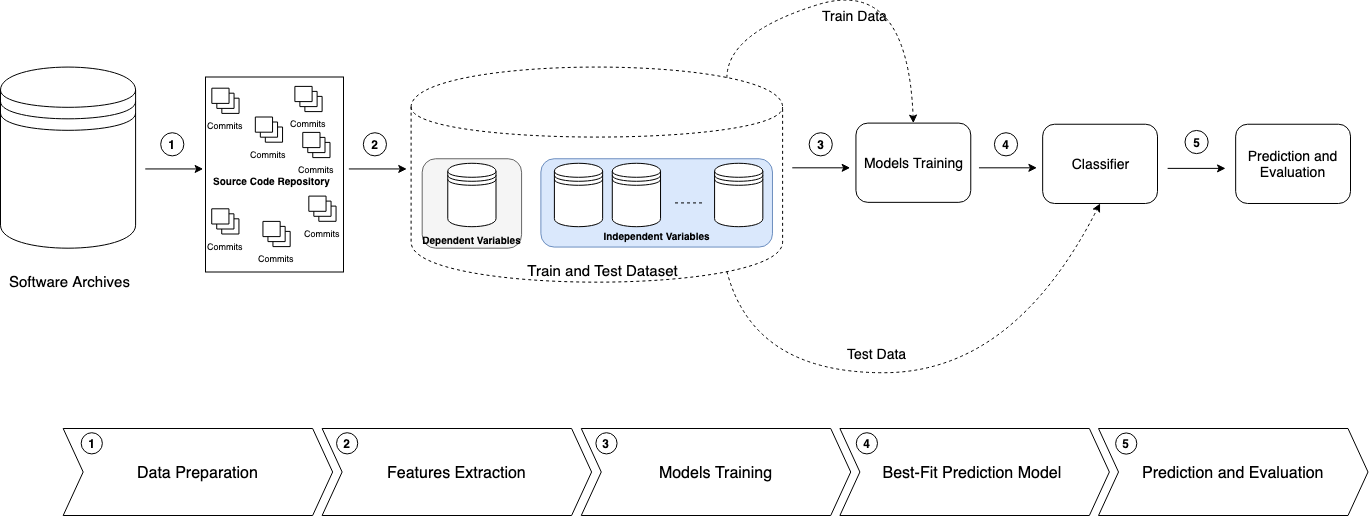}
\caption{Software Defect Prediction Process}
\label{fig:defect-prediction-process}
\end{figure*}

\begin{figure}[hbtp]
\centering
\includegraphics[scale=0.2]{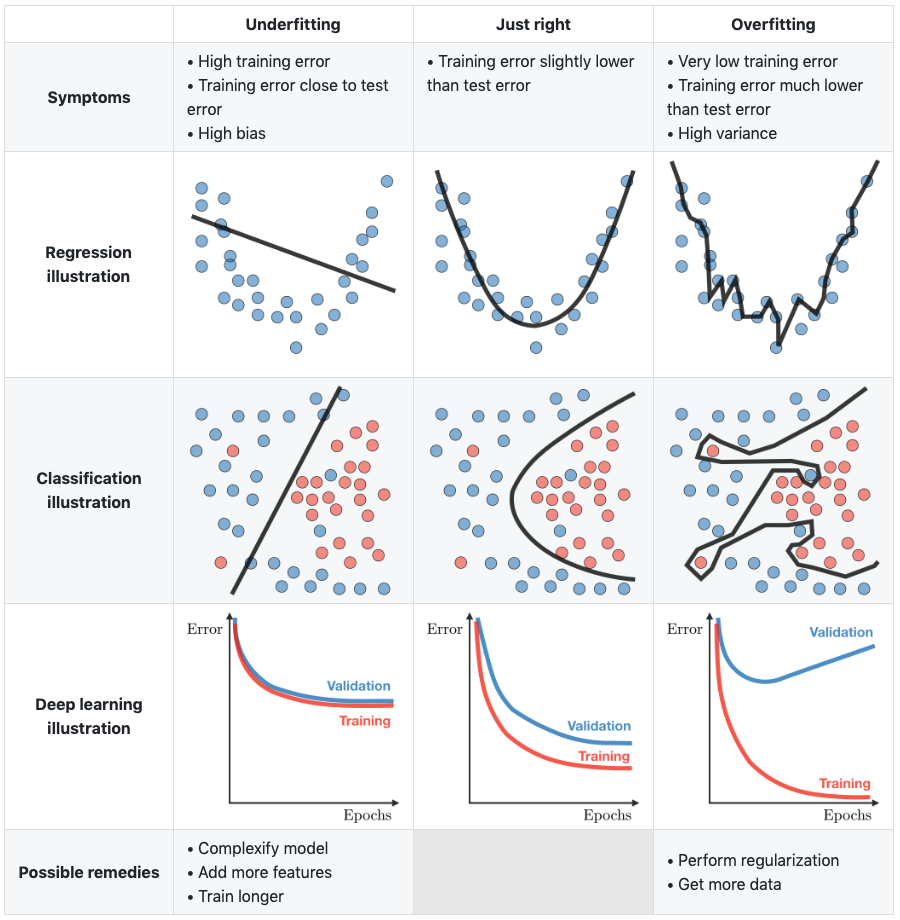}
\caption{Fitting Model Diagnostics \cite{afshine2018}}
\label{fig:illustration-fitting-models}
\end{figure}

\begin{figure}[hbtp]
\centering
\includegraphics[scale=0.3]{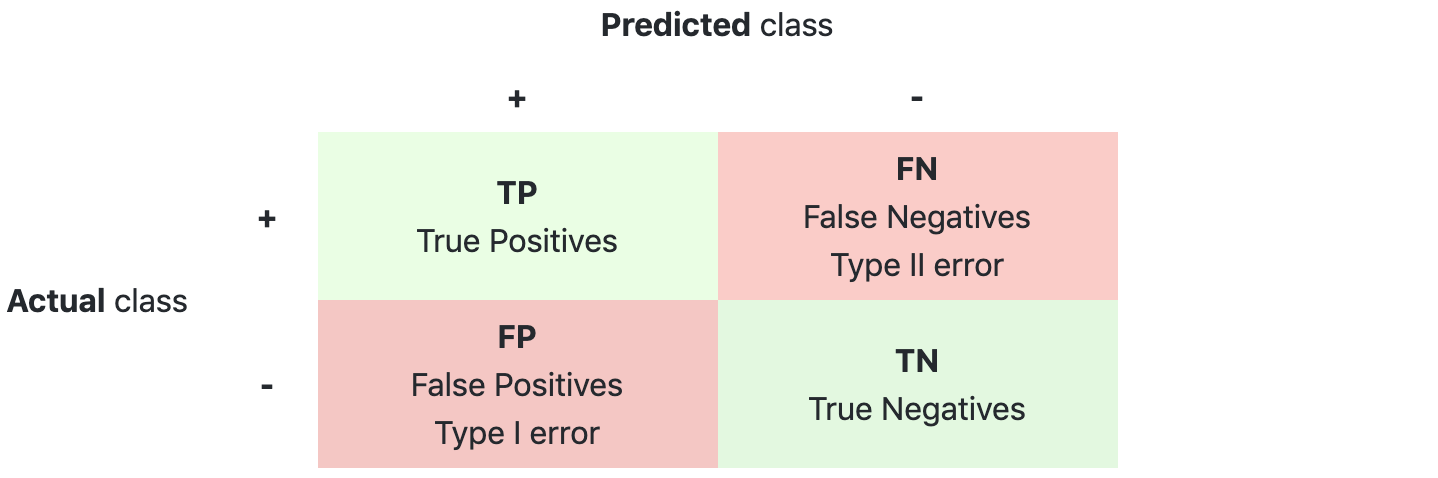}
\caption{Confusion Matrix}
\label{fig:confusion-matrix}
\end{figure}

As the process shows in Figure \ref{fig:defect-prediction-process}, the first step is to collect source code repositories from software archives.
The second step is to extract features from the source code repositories and the commits contained
therein. There
are many traditional features defined in past studies, which can be categorized into two kinds:
\begin{enumerate}
\item \textit{Metrics used as input for the common machine learning techniques:}\\
code metrics (e.g., McCabe features and CK features) and process metrics (e.g., change histories).
\item \textit{Metrics used as input for the deep learning techniques:}\\
an abstract representation of the source code (e.g., AST, control flow, data flow, etc...)
\end{enumerate}
The extracted features represent the train and test dataset. 
To select the best-fit defect prediction model, the most commonly used method is called $k$-fold cross-validation that splits the training data into $k$ groups to validate the model on one group while training the model on the $k - 1$ other groups, all of this $k$ times. The error is then averaged over the $k$ runs
and is named cross-validation error.
The diagnostics of the model is based on these features:
\textit{(1) Bias}: the bias of a model is the difference between the expected prediction and the correct model that we try to predict for given data points.
\textit{(2) Variance}: the variance of a model is the variability of the model prediction for given data points.
\textit{(3) Bias/variance tradeoff}: the simpler the model, the higher the bias, and the more complex the model, the higher the variance.
Figure~\ref{fig:illustration-fitting-models} shows a brief summary of how underfitting, overfitting and
a suitable fit looks like for the three commonly used techniques regression, classification and deep learning.
Once the model has been chosen, it is trained on the entire dataset and tested on the test dataset.
Most defect prediction approaches take defect prediction as a binary classification problem.
After fitting the models, the test data is fed into the trained classifier (the best-fit prediction model), which can predict whether the files are buggy or clean.
%that could be solved by different classification algorithms, e.g., Support Vector Machine
%(SVM), Naive Bayes (NB), Decision Tree (DT), Neural Network (NN)
%Such approaches simply classify source code changes into two categories: buggy or clean.
Afterwards, in order to assess the performance of the selected model, quality metrics are computed. To have a more complete picture when assessing the performance of a model, a confusion matrix is used. It is defined as shown in Figure~\ref{fig:confusion-matrix}.
We summarize the metrics for the performance of classification models in Table~\ref{tab:metrics-to-assess-performance-of-classification-models}.

\begin{table}[]
\centering
\caption{Common metrics used to assess the performance of classification models}
\begin{tabular}{|c|c|l|} \hline
\textbf{Metric} & \multicolumn{1}{l}{\textbf{Formula}} & \textbf{Interpretation} \\ \hline
Accuracy & $\frac{TP+TN}{TP+TN+FP+FN}$ & \begin{tabular}[c]{@{}l@{}}Overall performance \\ of model\end{tabular} \\ \hline
Precision & $ \frac{TP}{TP+FP}$ & \begin{tabular}[c]{@{}l@{}}How accurate the positive \\  predictions are\end{tabular} \\ \hline
\begin{tabular}[c]{@{}c@{}}Recall \end{tabular} & $ \frac{TP}{TP+FN}$ & \begin{tabular}[c]{@{}l@{}}Coverage of actual \\ positive sample\end{tabular} \\ \hline
F1 score & $ \frac{2TP}{2TP+FP+FN}$ & \begin{tabular}[c]{@{}l@{}}Hybrid metric useful for \\ unbalanced classes\end{tabular} \\ \hline
\end{tabular}
%\caption{Common metrics used to assess the performance of classification models}
\label{tab:metrics-to-assess-performance-of-classification-models}
\end{table}

\subsection{Within-Project Defect Prediction}
\textit{Within-project defect prediction}
%means that both the training data and test data come from the same project.
uses training data and test data that are from the same project. Many machine learning algorithms have been adopted for within-project defect prediction, including Support Vector Machines (SVM) \cite{elish2008}, Bayesian Belief Networks \cite{amasaki2003}, Naive Bayes (NB) \cite{taowang2010}, Decision Trees (DT) \cite{gayatri2010}, \cite{khoshgoftaar2002}, \cite{wang2012}, Neural Networks (NN) \cite{elhampaikari2012}, or Dictionary Learning \cite{jing2014}. 
Elish et al. \cite{elish2008} evaluated the feasibility of SVM in predicting defect-prone software modules, and they compared SVM against eight statistical and machine learning models on four NASA datasets. Their results showed that SVM is generally better than, or at least competitive with other models, e.g., Logistic Regression, Bayesian techniques, etc. 
Amasaki et al. \cite{amasaki2003} used a Bayesian Belief Network to predict the final quality of a software product. They evaluated their approach on a closed project, and the results showed that their proposed method can predict bugs that the Software Reliability Growth Model (SRGM) cannot handle. 
Wang et al. \cite{wang2012} and Khoshgoftaar et al. \cite{khoshgoftaar2002} examined the performance of tree-based machine learning algorithms on defect prediction. Their results indicate that tree-based algorithms can generate good predictions. 
Tao et al. \cite{taowang2010} proposed a Naive Bayes based defect prediction model, and they evaluated the proposed approach on $11$ datasets from the PROMISE defect data repository. Their experimental results showed that the Naive Bayes based defect prediction models could achieve better performance than $J48$ (decision tree) based prediction models. 
Jing et al. \cite{jing2014} introduced the dictionary learning technique to defect prediction. Their cost-sensitive dictionary learning based approach could significantly improve defect prediction in their experiments.
Wang et al.  \cite{wang2016} used a Deep Belief Network (DBN) to generate semantic features for file-level defect prediction tasks.
%%%%%% DEEP LEANING Study%%%%%%
In Wang et al.'s work \cite{wang2016}, to evaluate the performance of DBN-based semantic features 
as well as traditional features, they built prediction models by using three typical machine learning algorithms, i.e., ADTree, Naive Bayes, and Logistic Regression. Their experimental results show that the learned DBN-based semantic features consistently outperform the traditional defect prediction features on these machine learning classifiers.
Most of the above approaches are designed for file-level defect prediction. For change-level defect prediction, Mockus and Weiss \cite{mockus2000} and Kamei et al. \cite{kamei2013} predicted the risk of a software change by using change measures, e.g., the number of subsystems touched, the number of files modified, the number of added lines, and the number of modification requests. 
Kim et al. \cite{kim2008} used the identifiers in added and deleted source code and the words in change logs to classify changes as being fault-prone or not fault-prone.
Jiang et al. \cite{jiang2013} and Xia et al. \cite{xia2016} built separate prediction models with characteristic features and meta features for each developer to predict software defects in changes.
Tan et al. \cite{tan2015} improved change classification techniques and proposed online defect prediction models for imbalanced data. Their approach uses time sensitive change classification to address the incorrect evaluation introduced by cross-validation. 
McIntosh et al. \cite{mcintosh2018} studied the performance of change-level defect prediction as software systems evolve. 
Change classification can also predict whether a commit is buggy or not \cite{perl2015}, \cite{prechelt2014}, \cite{pradel2019}.
In Wang et al.'s work \cite{wang2016}, they also compare the DBN-based semantic features with the widely used change-level defect prediction features, and ther results suggest that the DBN-based semantic features can also outperform change-level features.

However, sufficient defect data is often unavailable for many projects and companies. This raises the need for cross-project bug localization, i.e., the use of data from one project to help locate bugs in another project.

\subsection{Cross-Project Defect Prediction}
Due to the lack of data, it is often difficult to build accurate models for new projects. 
%To address this issue, cross-project defect prediction (CPDP) models are trained by using data from other projects. 
Recently, more and more papers studied the \emph{cross-project defect prediction} problem, where the training data and test data come from different projects. 

Some studies (\cite{kitchenham2007}, \cite{menzies2010}, \cite{zimmermann2009}) have been done on evaluating cross-project defect prediction against within-project defect prediction and show that cross-project defect prediction is still a challenging problem.
He et al. \cite{he2013} showed the feasibility to find the best cross-project models among all available models to predict defects on specific projects. Turhan et al. \cite{turhan2009} proposed a nearest-neighbor filter to improve cross-project defect prediction. 
Zimmermann et al. \cite{zimmermann2009} evaluated the performance of cross-project defect prediction on $12$ projects and their $622$ combinations. They found that the defect prediction models at that time could not adapt well to cross-project defect prediction. 
%Premraj et al. \cite{premraj2011} compared network and code metrics for defect prediction, and further built six cross-project defect prediction models using those metrics sets. Their results confirmed that cross-project defect prediction is a challenging problem.
Li et al. \cite{li2017} proposed defect prediction via convolutional neural networks (DP-CNN). Their work differs from the above-mentioned approaches in that they utilize deep learning technique (i.e., CNN) to automatically generate discriminative features from source code, rather than manually designing features which can capture semantic and structural information of programs. Their features lead to more accurate predictions.
The state-of-the-art cross-project defect prediction is proposed by Nam et al. \cite{nam2013}, who adopted a state-of-the-art transfer learning technique called Transfer Component Analysis (TCA). They further improved TCA as TCA$+$ by optimizing TCA's normalization process. They evaluated TCA$+$ on eight open-source projects, and the results show that their approach significantly improves cross-project defect prediction. 
Xia et al. \cite{xia2016} proposed HYDRA, which leverages a genetic algorithm and ensemble learning (EL) to improve cross-project defect prediction. HYDRA requires massive training data and a portion ($5\%$) of labeled data from test data to build and train
the prediction models.
TCA$+$ \cite{nam2013} and HYDRA \cite{xia2016} are the two state-of-the-art techniques for cross-project defect prediction. However, in Wang et al.'s work \cite{songwang2018}, they only use TCA$+$ as baseline for cross-project defect prediction. This is because HYDRA requires that the developers manually inspect and label $5\%$ of the test data, while in real-world practice, it is very expensive to obtain labeled data from software projects, which requires the developers' manually inspection, and the ground truth might not be guaranteed. 
Most of the above existing cross-project approaches are examined for file-level defect prediction only. 
Recently, Kamei et al.  \cite{kamei2016} empirically studied the feasibility of change level defect prediction in a cross-project context.
Wang et al. \cite{songwang2018} examines the performance of Deep Belief Network (DBN)-based semantic features on change-level cross-project defect prediction tasks.
The main differences between this and existing approaches for within-project defect prediction and cross-project defect prediction are as follows. First, existing approaches to defect prediction are based on manually encoded traditional features which are not sensitive to the programs' semantic information, while Wang et al.'s approach automatically learns the semantic features using a DBN and uses these features to perform defect prediction tasks. 
Second, since Wang et al.'s method requires only the source code of the training and test projects, it is suitable for both within-project and cross-project defect prediction.
The semantic features can capture the common characteristics of defects, which implies that the semantic features trained from one project can be used to predict a different project, and thus is applicable in cross-project defect prediction. 

Deep learning-based approaches require only the source code of the training and test projects, and 
are therefore suitable for both within-project and cross-project defect prediction.
In the next session, we explain, based on recent research, how effective and accurate
fault-prediction models developed using deep learning techniques are.

\section{Deep Learning in Software Defect Prediction}

Recently, deep learning algorithms have been adopted to improve research tasks in software engineering. The most popular deep learning techniques are: Deep Belief Networks (DBN), Recurrent Neural Networks, Convolutional Neural Networks and Long Short Term Memory (LSTM), see Table~\ref{tab:dl-ml-techniques-summary}.
Yang et al. \cite{yang2015} propose an approach that leverages deep learning to generate new features from existing ones and then use these new features to build defect prediction models. Their work was motivated by the weaknesses of logistic regression (LR), which is that LR cannot combine features to generate new features. They used a Deep Belief Network (DBN) to generate features from $14$ traditional change level features, including the following: number of modified subsystems, modified directories, modified files, code added, code deleted, lines of code before/after the change, files before and after the change, and several features related to developers' experience \cite{yang2015}.
The work of Wang et al. \cite{songwang2018} differs from the above study mainly in three aspects. First, they use a DBN to learn semantic features directly from source code, while Yang et al. use relations among existing features. Since the existing features cannot distinguish between many semantic code differences, the combination of these features would still fail to capture semantic code differences. 
For example, if two changes add the same line at different locations in the same file, the traditional features cannot distinguish between the two changes. Thus, the generated new features, which are combinations of the traditional features, would also fail to distinguish between the two changes.
%There also many existing studies that leverage deep learning techniques to address other problems in software
%engineering \cite{gu2018}, \cite{gu2016}, \cite{guo2017}, \cite{wang2017}, \cite{lam2015}, \cite{li2017}, \cite{mou2016}, \cite{cakir2018}, \cite{raychev2014}, \cite{white2015}, \cite{xu2016}. Mou et al. \cite{mou2016} used deep learning to model programs and showed that deep learning can capture the programs' structural information. 

\begin{table*}[]
\centering
\caption{Common machine learning and deep learning techniques used in software defect prediction}
\scalebox{0.82}{
\begin{tabular}{|l|l|l|l|l|}
\hline
\textbf{Techniques} & \textbf{Definition} & \textbf{Advantages} & \textbf{Drawbacks} & \textbf{Ref.}\\ \hline
RNN & \begin{tabular}[c]{@{}l@{}}RNNs are called recurrent because \\ they perform the same task for \\ every element of a sequence, \\ with the output being depended on \\ the previous computations.\end{tabular} & \begin{tabular}[c]{@{}l@{}}- Possibility of processing \\ input of any length\\ - Model size not increasing \\ with size of the input\\ - Computation takes into \\ account historical information\end{tabular} & \begin{tabular}[c]{@{}l@{}}- Slow computation \\ - Difficulty of accessing \\ information from a \\ long time ago\\ - Cannot consider any future \\ input for the current state\end{tabular} & \begin{tabular}[c]{@{}l@{}} \cite{jinyong2018} \end{tabular}\\ \hline
LSTM & \begin{tabular}[c]{@{}l@{}}A long short-term memory (LSTM) \\ network is a type of RNN model \\ that avoids the vanishing gradient \\ problem by adding 'forget' gates.\end{tabular} & \begin{tabular}[c]{@{}l@{}}- Remembering information \\ for a long periods of time\end{tabular} & \begin{tabular}[c]{@{}l@{}}- It takes longer to train\\ - It requires more memory \\ to train\end{tabular}& \begin{tabular}[c]{@{}l@{}} \cite{dam2019}, \cite{dam2018} \end{tabular} \\ \hline
CNN & \begin{tabular}[c]{@{}l@{}}CNN is a class of deep neural network, \\ it uses convolution in place of general \\ matrix multiplication in at least one \\ of their layers.\end{tabular} & \begin{tabular}[c]{@{}l@{}}- It automatically detects the \\ important features without any\\  human supervision.\end{tabular} & \begin{tabular}[c]{@{}l@{}}- need a lot of training data.\\ - High computational cost.\end{tabular} & \begin{tabular}[c]{@{}l@{}} \cite{li2017}, \cite{mou2016}, \cite{phan2018} \end{tabular} \\ \hline
\begin{tabular}[c]{@{}l@{}}Stacked\\ Auto-Encoder\end{tabular} & \begin{tabular}[c]{@{}l@{}}A stacked autoencoder is a neural \\network consist several layers of sparse \\autoencoders where output of each hidden \\ layer is connected to the input of the \\successive hidden layer.\end{tabular} & \begin{tabular}[c]{@{}l@{}}- Possible use of pre-trained layers \\ from another model, to apply \\ transfer learning\\ - It does not require labeled inputs \\ to enable learning\end{tabular} & \begin{tabular}[c]{@{}l@{}}- Computationally expensive \\ to train\\ - Extremely uninterpretable\\ - The underlying math is more \\ complicated\\ - Prone to overfitting, though \\this can be mitigated \\via regularization\end{tabular} & \begin{tabular}[c]{@{}l@{}} \cite{manjula2019}, \cite{tong2017} \end{tabular} \\ \hline
DBN & \begin{tabular}[c]{@{}l@{}}DBN is an unsupervised probabilistic \\deep learning algorithm.\end{tabular} & \begin{tabular}[c]{@{}l@{}}- Only needs a small labeled dataset\\ - It is a solution to the \\ vanishing gradient problem\end{tabular} & \begin{tabular}[c]{@{}l@{}}- It overlooks the structural \\ information of programs\end{tabular} & \begin{tabular}[c]{@{}l@{}} \cite{wang2016} \end{tabular} \\ \hline
\begin{tabular}[c]{@{}l@{}}Logistic \\ Regression\end{tabular} & \begin{tabular}[c]{@{}l@{}}LR is used to describe data and \\ to explain the relationship between \\ one dependent binary variable and \\ independent variables.\end{tabular} & \begin{tabular}[c]{@{}l@{}}- Easy to implement\\ - Very efficient to train\end{tabular} & \begin{tabular}[c]{@{}l@{}}- It cannot combine different \\features \\ to generate new features.\\ - It performs well only when \\input features and output \\labels are in linear relation\end{tabular} & \begin{tabular}[c]{@{}l@{}} \cite{kamei2013} \end{tabular} \\ \hline
SVM & \begin{tabular}[c]{@{}l@{}}SVM is a supervised learning model.\\ It can be used for both regression \\ and classification tasks.\end{tabular} & \begin{tabular}[c]{@{}l@{}}- Using different kernel function it \\ gives better prediction result\\ - Less computation power\end{tabular} & \begin{tabular}[c]{@{}l@{}}- Not suitable for large number\\ of software metrics\end{tabular} & \begin{tabular}[c]{@{}l@{}} \cite{elish2008} \end{tabular} \\ \hline
Decision Tree & \begin{tabular}[c]{@{}l@{}}DT is a decision support tool that uses a \\ tree-like graph or model of decisions \\ and their possible consequences.\end{tabular} & \begin{tabular}[c]{@{}l@{}}Tree based methods empower \\ predictive models with high \\ accuracy, stability and \\ ease of interpretation.\end{tabular} & \begin{tabular}[c]{@{}l@{}}- Construction of decision tree \\is complex\end{tabular} & \begin{tabular}[c]{@{}l@{}} \cite{gayatri2010}, \cite{khoshgoftaar2002}, \cite{wang2012} \end{tabular} \\ \hline
\end{tabular}}
%\caption{Common machine learning and deep learning techniques used in software defect prediction}
\label{tab:dl-ml-techniques-summary}
\end{table*}
How to explain deep learning results is still a challenging question in the AI community. To interpret deep learning models, Andrej et al. \cite{karpathy2015} used character level language models as an interpretable testbed to explain the representations and predictions of a Recurrent Neural Network (RNN). 
Their qualitative visualization experiments demonstrate that RNN models could learn powerful and often interpretable long-range interactions from real-world data. 
Radford et al. \cite{radford2017} focus on understanding the properties of representations learned by byte-level recurrent language models for sentiment analysis. Their work reveals that there exists a sentiment unit in the well-trained RNNs (for sentiment analysis) that has a direct influence on the generative process of the model. 
Specifically, simply fixing its value to be positive or negative can generate samples with the corresponding positive or negative sentiment. 
The above studies show that to some extent deep learning models are interpretable.
However, these two studies focused on interpreting RNNs on text analysis. Wang et al. \cite{songwang2018}  leverages a different deep learning model, Deep Belief Networks (DBN), to analyze the ASTs of source code. The DBN adopts different architectures and learning processes from RNNs. For example, an RNN (e.g., LSTM) can, in principle, use its memory cells to remember long-range information that can be used to interpret data it is currently processing, while a DBN does not have such memory cells. Thus, it is unknown whether DBN models share the same properties (w.r.t interpretability) as RNNs.
Many studies used a topic model \cite{blei2003} to extract semantic features for different tasks in software engineering (\cite{chen2012}, \cite{nguyen2011}, \cite{xie2012}). Nguyen et al. \cite{nguyen2011} leveraged a topic model to generate features from source code for within-project defect prediction. However, their topic model handles each source file as an unordered token sequence.
Thus, the generated features cannot capture structural information in a source file.
A just-in-time defect prediction technique was proposed by Kamei et al. which leverages the advantages of Logistic Regression (LR) \cite{kamei2013}. However, logistic regression has two weaknesses.
First, in logistic regression, the contribution of each feature is calculated independently, which means that LR cannot combine different features to generate new ones. For example, given two features $x$ and $y$, if $x \times y$ is a highly relevant feature, it is not enough to input only $x$ and $y$ because logistic regression cannot generate the new feature $x \times y$. Second, logistic regression performs well only when input features and output labels are in linear relation. Due to
these two weaknesses, the selection of input features becomes
crucial when using logistic regression. The bad selection of
features may result in a non-linear relation for output labels,
leading to bad training performance or even training failure.
This severe problem leads some studies to adopt Deep Belief Network
(DBN), which is one of the state-of-the-art deep learning
approaches. 
%\begin{figure*}[hbtp]
%\centering
%\includegraphics[scale=0.3]{DBN-approach}
%\caption{DBN approach to generating semantic features for file-level and change-level defect
%prediction}
%\label{fig:DBN-approach}
%\end{figure*}
The biggest advantage of DBN, as shown in Table~\ref{tab:dl-ml-techniques-summary}, over logistic
regression is that DBNs can generate a more expressive feature set from the initial feature set. 
%The generated feature set, which can include $x \times y$, $x^{y}$, and even more complicated nonlinear combinations of the initial features, is more powerful to expressthe nature of a problem. 
%If we input these generated features instead of the initial set of basic features, the above two weaknesses with logistic regression can be overcome.
We summarizes in Table~\ref{tab:dl-ml-techniques-summary} the most commonly used machine learning and deep learning techniques in software defect prediction. 
\begin{table*}[]
\centering
\caption{Test cases prioritization approaches using machine learning techniques}
%\scalebox{0.95}{
\begin{tabular}{|l|l|l|l|l|}
\hline
\textbf{Techniques} & \textbf{Used Metrics}                                                                                                                                                                                         & \textbf{Advantages} & \textbf{Drawbacks}                                                                                                         & \textbf{Ref.}          \\ \hline
Neural Networks   & \begin{tabular}[c]{@{}l@{}}- Duration of test cases, \\ - Last execution, \\ - History of failures\end{tabular} & \begin{tabular}[c]{@{}l@{}}Test cases priorization done \\ with continuous actions, \\ Adaptive\end{tabular} & \begin{tabular}[c]{@{}l@{}}Needs a long \\ history of tests, needs \\ more data processing if \\ the test history is long\end{tabular} & \cite{spieker2017} \\ \hline
Bayesian Network  & \begin{tabular}[c]{@{}l@{}}- Changes of the source code, \\ - Coverage degree, \\ - Fault-proneness of the software\end{tabular} &\begin{tabular}[c]{@{}l@{}} Hight value of APFD                                                                                                    \\(Average Percentage Faults Detected)\end{tabular} & \begin{tabular}[c]{@{}l@{}}Requires a reasonable \\ number of available faults\end{tabular}                                            & \cite{mirarab2007}   \\ \hline
Bayesian Network  & \begin{tabular}[c]{@{}l@{}}- Quality metrics, \\ - Code change, \\ - Coverage degree\end{tabular} & \begin{tabular}[c]{@{}l@{}}Improves performance of detecting \\ faults earlier\end{tabular} & Time-consuming & \cite{joachims2002}  \\ \hline
Bayesian Network  & \begin{tabular}[c]{@{}l@{}}- Similarity on code coverage, \\ - Probability of failure\end{tabular} & \begin{tabular}[c]{@{}l@{}}Achieves full coverage, \\ Faster fault detection\end{tabular} & Doesn't use real faults & \cite{zhao2015}      \\ \hline
Genetic algorithm & - Code coverage                                                                                                                                                                                      & Adaptive, Simple &                                                                                                                                       \begin{tabular}[c]{@{}l@{}}Do not scale well\\ with complexity\end{tabular} & \cite{konsaard2015}  \\ \hline
Genetic algorithm & \begin{tabular}[c]{@{}l@{}}- Code degree of statements, \\ - Multiple conditions\end{tabular} & \begin{tabular}[c]{@{}l@{}}Considers severities of faults, \\ time of execution and structural \\ coverage\end{tabular} & \begin{tabular}[c]{@{}l@{}}Do not scale well\\ with complexity\end{tabular} & \cite{ahmed2012}     \\ \hline
Genetic algorithm & \begin{tabular}[c]{@{}l@{}}- Fault proneness of software modules, \\ - Faults probability of each test case\end{tabular} & \begin{tabular}[c]{@{}l@{}}More efficient than the \\non-adaptive approach \end{tabular} & \begin{tabular}[c]{@{}l@{}}No optimization if there is \\ no time to execute test cases\end{tabular} & \cite{abele2014}     \\ \hline
SVM & \begin{tabular}[c]{@{}l@{}}- Coverage degree of requirements, \\ - Failure count, \\ - Failure age, \\ - Priority of failures,\\  - Cost of executing tests, \\ - Test case description\end{tabular} & \begin{tabular}[c]{@{}l@{}}Failure detection rate is increased \\ compared to a random order\end{tabular} & Based on human intuition                                                                                                           & \cite{lachmann2016}  \\ \hline
SVM & \begin{tabular}[c]{@{}l@{}}- Code coverage, \\ - Textual similarity between tests and changes, \\ - Test-failure history, \\ - Fault history, \\ - Test age\end{tabular} & \begin{tabular}[c]{@{}l@{}}\\ Convenient for industrial settings\end{tabular} & \begin{tabular}[c]{@{}l@{}}Integration of multiple\\ techniques\end{tabular} & \cite{busjaeger2016} \\ \hline
K-means & - Fault detection capability                                                                                                                                                                           & Good for large scale testing &                                                                                                                                      Dependent on initial values & \cite{chen2017}      \\ \hline
\end{tabular}
\label{tab:tcp-ml-techniques-summary}
\end{table*}

\section{Machine Learning-Driven Test Case Prioritization}

Test case prioritization aims to decrease the cost of regression testing by finding a test case ordering that maximizes the fault detection capability of the test suite. It seeks to find the optimal permutation of the sequence of test cases. It does not involve selection of test cases, and assumes that all the test cases may be executed in the order of the permutation it produces, but that testing may be terminated at some arbitrary point during the testing process. More formally, the prioritization problem is defined as follows \cite{yoo2012}:\\

%Test case prioritization aims to decrease the cost of regression testing by finding a test case ordering that maximizes the fault detection capability of the test suite, such that any faults in a program can be quickly identified by running a reduced number of tests. Definition 1 defines it as:
\textbf{Definition:}
\textit{test case prioritization requirements:}\\
$T,$ a test suite\\
$PT$, the set of permutations of $T$\\
$f$ , a function that gives a numerical score for $T'\in PT$\\
\textit{Problem:}\\
Find $T'$ such that
\begin{equation*}
(\forall T'') (T'' \in PT) (T'' \neq T') [f(T') \geq f(T'')].
\end{equation*}

In order to maximize the fault detection capability of the test suite, an appropriate $f$ function must be chosen to select the permutation $T'$ that finds all regressions as soon as possible. Since we cannot know about either the existence or location of faults prior to running the test suite, $f$ can only be a surrogate for actual fault detection. Most previous research on test case prioritization has tried to find an implementation of $f$ that most closely approximates fault detection, using a variety of strategies \cite{yoo2012}, \cite{gonzalez2011}.

\textit{Machine learning for software testing:} Machine learning algorithms are increasingly being considered in the context of software testing. Busjaeger and Xie \cite{busjaeger2016} use machine learning and multiple heuristic techniques to prioritize test cases in an industrial environment. Through combining different data sources and learning agnostic prioritization, this work makes an important step towards defining a general framework for automatically learning the prioritization of test cases. Spieker et al. \cite{spieker2017} offer a lightweight learning method that uses a single data source, namely the test case failure history based on reinforcement learning and artificial neural networks. Chen et al. \cite{chen2011} uses semi-supervised clustering for regres-sion test selection. 
The drawback of this approach can be a higher computational complexity. Some other approaches include active learning for test classification \cite{bowring2004}, the combination of machine learning and program slicing for regression test case prioritization \cite{wang2012}, learning agent-based test case prioritization \cite{abele2014}, or clustering approaches \cite{chaurasia2015}. Groce et al. \cite{groce2012} proposed reinforcement learning (RL) for automated testing of software APIs in combination with adaptation-based programming (ABP). The combination of RL and ABP selects successive calls to the API with the goal of increasing test coverage. Moreover, Reichstaller et al. \cite{reichstaller2016} use RL to generate test cases for risk-based interoperability tests. 
%Based on a model of the system under test, RL agents are trained to interact in an error-provoking way, i.e. they are encouraged to exploit possible interactions between components. 
%Other reseachers \cite{veanes2006} have used reinforcement learning for online formal testing of communication systems. For this purpose, they are using RL to reinforce the behavior of the testers when For this purpose, they are using RL to reinforce the behavior of the testers when modeling system and test cases as input-output marked transition systems. system and test cases as input-output labeled transition systems.
We summarize in Table~\ref{tab:tcp-ml-techniques-summary} the approaches with the common used techniques in test cases prioritization.
A number of machine learning techniques are used to prioritize test cases. The most commonly used techniques are Neural Networks, Bayesian Networks and genetic algorithms.
Spieker et al. \cite{spieker2017} show that neural networks are very good classifiers and are used to group test cases based on similar properties.
This simplifies the process of ordering test cases for the test cases priorization process. Bayesian network was used in numerous approaches \cite{zhao2015}, \cite{joachims2002}, \cite{mirarab2007}, are well adapted to calculate probabilities which are used to order test cases. In addition, other researchers \cite{abele2014}, \cite{ahmed2012}, \cite{konsaard2015} use genetic algorithms which order test cases that execute many iterations with three steps each: selection of some of the test cases based on a fitness function that uses some predefined characteristics for selection; crossover and mutation. Neural networks and Bayesian networks are very adaptable to the problem and very efficient in integrating many features.
Numerous other machine learning methods have also proven to be efficient in TCP. Support Vector Machines (SVM) were used in some approaches \cite{lachmann2016}, \cite{busjaeger2016}, SVM are also effective in integrating different attributes while K-nearest neighbor, logistic regression and k-means can be used in clustering test cases \cite{chen2017}. 
%Different machine learning techniques results can also be combined to create an ensemble learner. This approach seems to have good results with the challenge of how to combine the results effectively. We can conclude that in general machine learning techniques improve the performance for solving TCP problem with a drawback that some of these techniques require a big dataset for training.

%%%%%%%%%%%%%%%%
%%%%%%%%%%
%Test case prioritization is an effective and practical technique that helps to increase the rate of regression fault detection when software evolves. Numerous techniques have been reported in the literature on prioritizing test cases for regression testing. However, existing prioritization techniques implicitly assume that source or binary code is available when regression testing is performed, and therefore cannot be implemented when there is no program source or binary code to be analyzed. 

%\section{Experiments and Results}
%\subsection{Black-Box Approach Using Traditional Machine Learning Techniques}
%\paragraph{Metrics Used}
%\paragraph{Results}
%\subsection{White-Box Approach Using Deep Learning Techniques}
%\paragraph{Metrics Used}

\section{Conclusion}
With the ever-increasing scale and complexity of modern software, software reliability assurance has become a significant challenge. To enhance the reliability of software, we consider predicting potential code defects in software implementations a beneficial direction, which has the potential to dramatically reduce the workload of software maintenance. Specifically, we see the highest potential in a defect prediction framework which utilizes deep learning algorithms for automated feature generation from source code with the semantic and structural information preserved. 
Moreover, our survey corroborates the feasibility of machine learning techniques in the filed of test cases prioritization.
%It seems also promising to adapt deep learning to other software engineering tasks such as code completion and code clone detection.

\bibliographystyle{IEEEtran}
\bibliography{sample-base}

\end{document}